\documentclass[conference]{IEEEtran}
\IEEEoverridecommandlockouts
\newcommand\scalemath[2]{\scalebox{#1}{\mbox{\ensuremath{\displaystyle #2}}}}
\usepackage{cite}
\usepackage{amsmath,amssymb,amsfonts}
\usepackage{algorithmic}
\usepackage{graphicx}
\usepackage{textcomp}
\usepackage{xcolor}
\usepackage{comment}
\usepackage{hyperref}
\hypersetup{
    colorlinks=true,
    linkcolor=blue,
    urlcolor=cyan,
    }
\usepackage{multicol}
\usepackage{mwe}
\usepackage[switch]{lineno}

\DeclareRobustCommand{\IEEEauthorrefmark}[1]{\smash{\textsuperscript{\footnotesize #1}}}
    
\begin{document}

\title{Iterative Feature Boosting for Explainable \\Speech Emotion Recognition}

\author{\IEEEauthorblockN{Alaa Nfissi\IEEEauthorrefmark{1,2,4}\hspace{0.5cm}
Wassim Bouachir\IEEEauthorrefmark{1,4}\hspace{0.5cm}
Nizar Bouguila\IEEEauthorrefmark{2}\hspace{0.5cm}
Brian Mishara\IEEEauthorrefmark{3,4}} \\
\IEEEauthorblockA{\textit{\IEEEauthorrefmark{1} Data Science Laboratory, University of Québec (TÉLUQ), Montréal, Canada}}
\IEEEauthorblockA{\textit{\IEEEauthorrefmark{2} Concordia Institute for Information Systems Engineering, Concordia University, Montréal, Canada}}
\IEEEauthorblockA{\textit{\IEEEauthorrefmark{3} Psychology Department, University of Québec at Montréal, Montréal, Canada}}
\IEEEauthorblockA{\textit{\IEEEauthorrefmark{4} Centre for Research and Intervention on Suicide, Ethical Issues and End-of-Life Practices, Montréal, Canada}
}}

\maketitle
\begin{abstract}
In speech emotion recognition (SER), using predefined features without considering their practical importance may lead to high dimensional datasets, including redundant and irrelevant information. Consequently, high-dimensional learning often results in decreasing model accuracy while increasing computational complexity. Our work underlines the importance of carefully considering and analyzing features in order to build efficient SER systems. We present a new supervised SER method based on an efficient feature engineering approach. We pay particular attention to the explainability of results to evaluate feature relevance and refine feature sets. This is performed iteratively through feature evaluation loop, using Shapley values to boost feature selection and improve overall framework performance.
Our approach allows thus to balance the benefits between model performance and transparency. The proposed method outperforms human-level performance (HLP) and state-of-the-art machine learning methods in emotion recognition on the TESS dataset. The source code of this paper is publicly available at \href{https://github.com/alaaNfissi/Iterative-Feature-Boosting-for-Explainable-Speech-Emotion-Recognition}{Iterative-Feature-Boosting-for-Explainable-Speech-Emotion-Recognition}.
\end{abstract}

\begin{IEEEkeywords}
Feature selection, Supervised learning, Speech emotion recognition, Acoustic analysis, Explainable AI.
\end{IEEEkeywords}

\section{Introduction}
\label{intro}
Human emotions are complex and essential aspects of human behavior. They correspond to subjective experiences characterized by physiological, behavioral, and cognitive changes, and can be influenced by various factors, such as social interactions, and cultural considerations. Emotions play an important role in various aspects of human life. Therefore, emotion recognition and interpretation is considered as a key problem in psychology, with a wide range of AI applications, including human-computer interaction, affective computing, and mental disorder detection \cite{shaver1987emotion}.

To solve the speech emotion recognition (SER) problem, various machine learning algorithms have been explored, including support vector machines \cite{6512793}, hidden Markov models \cite{nwe2003speech}, and deep neural networks \cite{fayek2017evaluating}. These algorithms are trained on large datasets, and use various feature representations, such as spectral features, prosodic features, and spectral envelope features \cite{el2011survey}. Given the multitude of features and their several categories, one of the main challenges is to find the appropriate feature representation for a given SER task. In fact, the use of large predefined feature sets often leads to high-dimensional datasets \cite{al2015automatic}, making it difficult for the model to learn effectively from data, in addition to increasing computational complexity \cite{ashok2022handling}. 

In the literature, little attention has been paid to finding relevant features for SER as most of existing works used principal component analysis (PCA) for dimensionality reduction \cite{fewzee2012dimensionality}, while some other works relied on 1D convolutions through end-to-end deep learning models \cite{nfissi2022cnn}. 
In our work, we argue that the performance of SER systems heavily depends on the used features. We thus present a comprehensive approach, placing a special emphasis on the feature extraction and selection process.

The proposed framework comprises three main components : 1) a feature boosting module guided by the feedback loop of the third component to extract and select features, 2) a classification module using a supervised classification model, and 3) an explainability module where the contribution of features to the classification decision is evaluated using SHapley Additive exPlanations (SHAP) \cite{roth1988shapley}. This explainability component serves as a feedback mechanism at the end of each iteration, to continuously refine and boost the feature set in the first component.

Accordingly, our main contributions can be summarized as follows:
\begin{itemize}
    \item A new SER approach with an emphasis on feature selection through iterative feature boosting.
    \item Incorporation of model explainability and SHAP technique for identifying the most relevant features for the SER task, as well as for transparency purposes.
    \item An experimental evaluation of the proposed method by comparison to human-level performance and state-of-the-art algorithms.
    \item The source code of our framework for reproducibility and future research on SER.
\end{itemize}

\section{RELATED WORKS}

\begin{figure}[t]
\centering
\scalemath{0.5}{
\centerline{\includegraphics[width=1\textwidth]{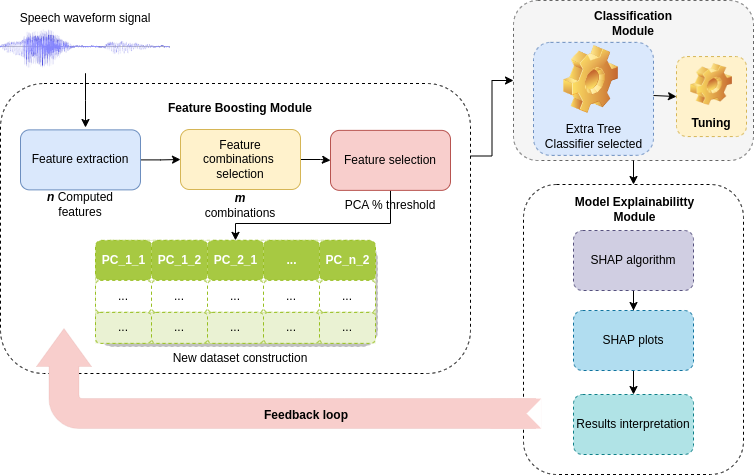}}
}
\caption{The proposed method}
\label{fig12}
\end{figure}

Modern SER methods are mostly based on supervised learning approaches. They can be broadly divided into two categories: traditional machine learning methods \cite{Iqbal2020MFCCAM} \cite{krishnan2021emotion}, and deep learning methods \cite{aggarwal2022two} \cite{praseetha2018deep} \cite{nfissi2022cnn}.
Several techniques related to feature extraction have been explored in previous works. They are mostly based on handcrafted features, which are designed by incorporating expert knowledge or domain-specific insights, and using traditional feature selection techniques. 
One of the major challenges is handling the high dimensionality of the data and ensuring that the used features are meaningful and relevant. This is because a large number of features are often extracted from signals without considering their practical importance and suitability for the emotion classification task. 

In \cite{Iqbal2020MFCCAM}, the authors propose to use Mel Frequency Cepstral Coefficients (MFCC) for feature extraction and Support Vector Machines (SVM) for classification. MFCCs are calculated through a series of steps including sampling and quantization, windowing, discrete Fourier transform, and a Mel filter bank. The resulting MFCCs are then used for emotion classification using an SVM classifier. A sensitivity analysis is also conducted to evaluate the impact of different feature combinations on the classification performance. 

In \cite{aggarwal2022two}, the authors present two-way approach for SER. The first involves the direct extraction of features from the audio dataset using a combination of mel-scale related features. Then, PCA is used to reduce data dimensionality and eliminate correlated variables. 
The resulting features are then fed into a deep neural network (DNN) for classification. The authors observed that PCA allowed to significantly reduce overfitting, which in turn leads to a more effective training of the DNN. Their second approach is based on using 2D representation of spectrograms considered as images for classification, which is then carried out by the VGG16 CNN \cite{simonyan2014very} model retrained on Mel-Spectrogram images to classify emotions. 

In \cite{song2020speech}, the authors present the robust discriminative sparse regression (RDSR) approach to deal with feature selection and emotion classification in a joint learning framework. The RDSR algorithm is designed to select the most discriminative feature subset from the original high-dimensional feature set. It uses sparse regression to improve model robustness to outliers and noise, and introduces a feature selection regularization constraint to select the most relevant features. 

In \cite{8785867}, an SER model based on continuous hidden Markov model (CHMM) was proposed, as the random generation of states in HMMs allows for statistical modeling of the sequential nature of the data. The model extracts 33-dimensional feature parameters based on temporal sequence and uses PCA to reduce the dimensionality of initial feature set. The experimental results showed that the PCA-CHMM model improves emotion recognition performance compared to the standard HMM model using the entire feature set.

In \cite{nfissi2022cnn}, the authors propose a hybrid end-to-end deep learning model for feature extraction and emotion classification. The proposed model consists of two main components: one-dimensional convolutional neural network (1D-CNN) and Gated Recurrent Unit (GRU). The 1D-CNN serves for spatial features extraction through convolutional layers, while the GRU component captures time-distributed features due to memory and gate principles.

Previous research has shown that extracting efficient acoustic characteristics is crucial for capturing various emotional aspects of speech in SER. However, existing works mostly relied on predefined features, without fully investigating their relevance for SER, and how they can be used to improve performance. Some deep learning-based works attempted to address the feature selection issue implicitly through 1D convolutions,  while other supervised learning methods aimed to handle the high dimensionality problem by merely applying PCA to compute principal components. We present a new approach for supervised SER with a focus on feature importance and model explainability over the entire framework.

\section{PROPOSED METHOD}
\label{prop_meth}
\subsection{Motivation and overview}

Both the quality and quantity of used features can significantly impact the performance of emotion recognition. Using too many irrelevant or redundant features can lead to overfitting and lack of generalization, while using few or insufficient features can result in underfitting and poor performance. Therefore, it is important to carefully select the features that are most useful for emotion recognition, by considering the voice signal characteristics that are most indicative of emotions. 

Our method is based on domain-specific voice features and comprises three main components. A feature boosting module  is firstly used to compute a preliminary feature set assumed to be useful for emotion recognition, these features are iteratively refined afterwards via a feedback mechanism. Then, we identify optimal feature combinations from the resulting feature set and reduce dimensionality. Secondly, a classification module formulates the SER task as a supervised classification problem. Classification decisions are then analyzed by an explainability module, which utilizes Shapley values to evaluate the importance of features in the classification decision, thus, to provide insights into the feature boosting module. By incorporating explainability into the SER process, we aim to enhance the performance of the model with a better understanding of the contribution of each feature to the final decision. The obtained explainability results are then used iteratively via a feedback mechanism to further boost the feature selection process in the first module. This is achieved through a feedback loop at the end of each iteration. 

\subsection{Method}

The proposed method is illustrated in Fig. \ref{fig12}. In the feature boosting module, we compute a preliminary feature set representation including pitch, energy, and rhythm related characteristics, which we assume being meaningful for SER. We also calculate related statistical characteristics including the mean, median, standard deviation, minimum and maximum \cite{yildirim2004acoustic}, resulting in a set of $n$ initial features. To further improve the performance and interpretability of our technique, we select $m$ optimal distinct combinations of $p$ features from the dataset according to following steps. We use PCA to reduce data dimensionality of each combination $i$ and remove noise by transforming it using an eigenvector matrix $(A_i)$ and a corresponding eigenvalues vector ($\lambda_i$). Each column of the eigenvector matrix represents a principal component $(PC_{ij})$ capturing specific data information and determining the dimension $(r_i)$ of the reduced subset (see eq. \ref{eq1} and \ref{eq2}).
\begin{multicols}{2}
  \begin{equation}
 A_i = 
\scalemath{0.72}{
\begin{bmatrix}
a_{i11} & a_{i12} & .. &	a_{i1r_i} \\	
.. & .. & .. & .. \\	
.. & .. & .. & ..  \\	
a_{ip1} & ..  & .. & a_{ipr_i} \\	
\end{bmatrix}
}
\label{eq1}
\end{equation}

\begin{equation}
\lambda_i = 
\scalemath{0.72}{
\begin{bmatrix}
    \lambda_{i1} \\ 
    .. \\ 
    .. \\
  \lambda_{ip}
\end{bmatrix}
}
\label{eq2}
\end{equation}
\end{multicols}
The percentage of variance $(EV_{ij})$ explained by $j^{th}$ principal component of the $i^{th}$ combination $(PC_{ij})$ can be evaluated using eq. \ref{eq3}:
\begin{equation}
EV_{ij} = \frac{\lambda_{ij}}{\sum_{j=1}^{p}\lambda_{ij}} \times 100 \label{eq3}
\end{equation}
where $\lambda_{ij}$ represents the eigenvalue and the amount of variance of the $j^{th}$ principal component of the $i^{th}$ combination $(PC_{ij})$. The expression of $(PC_{ij})$ is given by eq. \ref{eq4}, where $X_{ik}$ is the $k^{th}$ feature of the $i^{th}$ combination:
\begin{equation}
PC_{ij} = (a_{i1j})X_{i1}  + ... + (a_{ipj})X_{ip} \label{eq4}
\end{equation}
Therefore, we can determine which features contribute the most to each principal component. This is done in order to identify the best combination of features representing information in our dataset. We use a threshold $\alpha$ on the sum of the first $c$ explained variances as a criterion to determine feature combinations that are most informative. The resulting feature set consists of the principal components of each selected combination. In this way, we eliminate redundant and less informative features, and use only the most relevant ones. This above detailed process is refined iteratively through the feedback loop of our architecture's explainability module, which we will describe later in this section.

In the classification module we compare the performance of $M$ candidate classification models on the resulting features, these models are detailed in section \ref{exp_set}. Our choice of classification models aligns with those commonly used classifiers in the SER literature. The selected machine learning models have been widely used in various domains for SER and have been known to achieve good results \cite{el2011survey}. The process of comparing the performance of $M$ different classification models on the initial dataset and the resulting feature set helps us determine how well our approach performs, to consequently select the appropriate model. Beyond SER, the framework could be applied to a wide range of classification problems, where other candidate classification models could be evaluated.

\begin{figure}[t]
\centering
\begin{minipage}[b]{.22\textwidth}
\includegraphics[width=\textwidth]{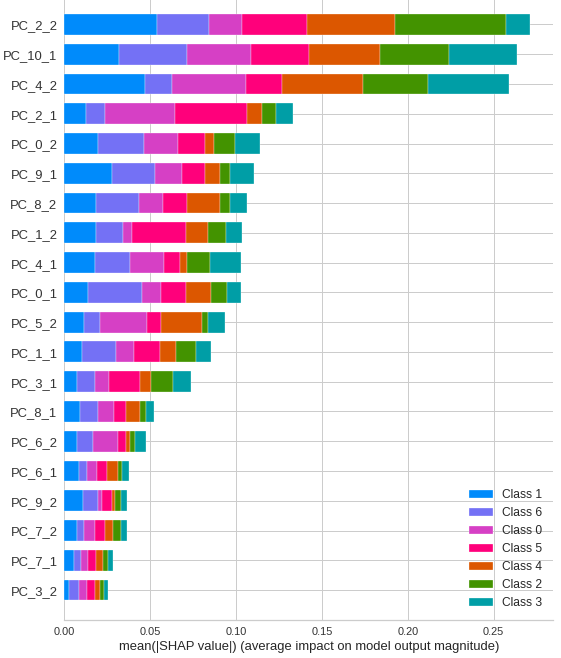}
\caption{Boosted features importance}\label{fig7}
\end{minipage}\qquad
\begin{minipage}[b]{.229\textwidth}
\includegraphics[width=\textwidth]{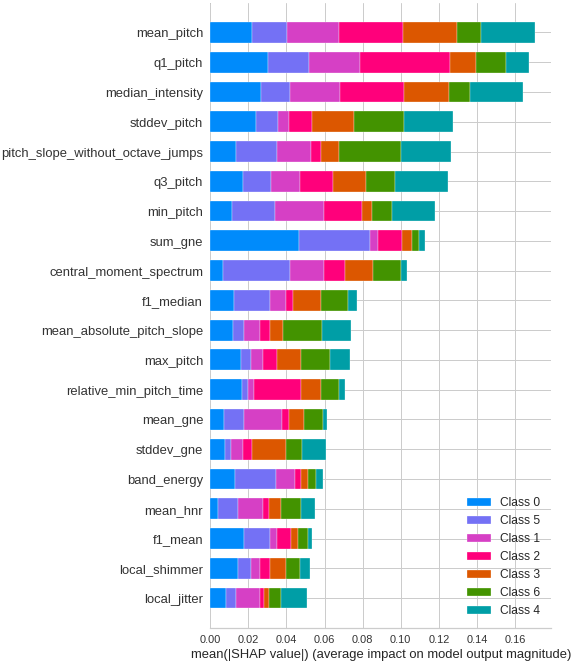}
\caption{Initial features importance}\label{fig8}
\end{minipage}\\
\scriptsize{\textit{Class-wise feature importance - Extra Trees classifier: \scriptsize The values on the x-axis are the mean of absolute values of Shapley values, which represent the average impact on the model's output magnitude, where a higher value indicates a more important feature. The features on the y-axis in Fig. \ref{fig7} are represented as principal components, which are the result of applying PCA to the optimal combinations of selected features using the explained variance threshold. In Fig. \ref{fig8}, the y-axis represents the initial feature importance.}}
\end{figure}

In the explainability module we incorporate explainable artificial intelligence (XAI) \cite{samek2019explainable} capabilities into the SER system. Thus, we create a system that is transparent and understandable in terms of prediction and decision making. 
To achieve this, we use the Shapley explanation values to explain the model's predictions. Shapley values allow to understand the contribution of each $PC_{ij}$ in the resulting feature set to a model's prediction, which enables us to identify which combination's principal components are most important for the emotion recognition. We define the contribution of each $PC_{ij}$ in the resulting feature set denoted $\phi_{PC_{ij}}$ as:
\begin{equation}
   \phi_{PC_{ij}} = \sum_{S\subseteq {11,\dots,mJ_{i}}\backslash{ij}}\frac{1}{|S|}\sum_{s\subseteq S}(-1)^{|s|+1}f_{ij}(z_{s\cup{ij}})-f(z_{s}) 
   \label{eq5}
\end{equation}
where $m$ is the number of combinations, $J_{i}$ is the number of principal components used from each combination $i$ with $J_{i}\geq1$, $S$ is a subset of principal components indices, $z$ is the input vector, and $f(z)$ is the output of the classification model for input $z$. The first summation term $(f_{ij}(z_{s\cup{ij}}))$ computes the expected output of the model when $PC_{ij}$ is included in the subset, while the second term $(f(z_{s}))$ computes the expected output of the model when $PC_{ij}$ is discarded from $S$. As a result, the difference between the two terms represents the contribution of $PC_{ij}$ to the output, which we use to conduct a feature importance analysis. This allows to get insight into how the model works and what factors are most important. 

The process involves an iterative feedback loop where we use the explainability module to determine the most relevant principal components that capture the essential information for emotion recognition and eliminate less important ones. We then incorporate the contribution of the initial features to each of the relevant principal components to identify the most participating features to the principal components, and thus, to the classification decision process. We eliminate features that do not contribute significantly to the model's performance and iterate the process until convergence.

\section{EXPERIMENTS}

\subsection{Dataset and experimental setup}
We use the Toronto Emotional Speech Set (\textbf{TESS}) \cite{dupuis2010toronto} dataset including 2800 audio recordings of two participants expressing 200 target phrases in different emotional states. The emotions included in the dataset are anger, disgust, pleasant surprise, fear, sadness, happiness, and neutral, where each emotion is represented by 400 recordings.

\label{exp_set}
We first set the sampling rate of the audio data to 16 KHz using a mono-channel format. This ensures that audio signals are properly processed and analyzed by our system, as most SER algorithms require specific sampling rates and number of channels for each audio signal. After going through the feature extraction and selection process, we use stratified random sampling \cite{aoyama1954study} to divide both original dataset and boosted features dataset into three homogeneous groups (or strata): training, validation, and testing. We keep 10\% of the data as unseen to be used for testing, 80\% for training, and 10\% for validation. This ensures that the distribution of classes is maintained across all subsets. Then, we use 10-fold cross-validation \cite{refaeilzadeh2009cross} to train $M=7$ machine learning models: Extra Trees (ET), Light Gradient Boosting Machine (LGBM), Random Forest (RF), Quadratic Discriminant Analysis (QDA), Gradient Boosting Classifier (GBC), Linear Discriminant Analysis (LDA), and Decision Tree (DT), on both datasets to select the optimal model for each. By using cross-validation, our performance evaluation should be less sensitive to data partitioning.

In order to improve the performance of our best-performing machine learning models, we use the grid search technique, which involves exhaustively searching through a specified parameter space to find the best combination of hyperparameters for a given model \cite{yang2020hyperparameter}. In this way, we are able to fine-tune the model by adjusting its hyperparameters to increase robustness. We thus find the optimal set of hyperparameters producing the highest performance on the validation dataset. 

We then assess the performance of the final models on the testing set. The testing performance is an indicator of how well the model would perform on unseen data without overfitting the training set. Finally, we use the SHAP approach in our explainability module to evaluate the feature importance in the predictions of the optimal model. This allows us to understand how the model is making its predictions and to identify which features are most important for determining emotions. For performance evaluation, we use accuracy, recall, precision, and F1-Score metrics.

\subsection{Model explainability}

One key aspect of our approach is the use of feature boosting module to select the optimal combinations of features that best capture the variance and information in our dataset. To demonstrate this, we show importance values in figures \ref{fig7} and \ref{fig8}. We can observe the contribution of each feature to the predicted emotion class. The principal components in Fig. \ref{fig7} are labeled as $PC_{\{combination\_index\}\{PC\_index\}}$, i.e. $PC_{ij}$ refers to the $j^{th}$ principal component of the $i^{th}$ feature combination. Additionally, it can also help in understanding which principal components are not important and can be removed without relatively affecting the performance of the model. By examining Fig. \ref{fig7}, we can determine the principal components that have the highest impact on the output of our model. Through this analysis, we can also identify the feature combination that correspond to each principal component and the contribution of each feature to a principal component. 
This leads us ultimately to determine the most relevant features for our SER decision (see Fig. \ref{fig8}). Thus, we can identify the features that contribute the most to each of the important principal components, this information is valuable because it allows us to refine our SER system by eliminating less relevant features and improving the accuracy of the classification decision in the next iteration.

We set a threshold $\alpha=0.8$ on the cumulative explained variance by the first two principal components of each of the feature combinations. This means that we select only the combinations that explain at least 80\% of the data variance. Thus, we obtain the best 19 combinations representing the data.
As an example, we have found a combination of features that are particularly informative. This feature set captures 84.56\% of the total explained variance in the first two principal components and 98.22\% in the first four. 

\begin{figure}[t]
\centerline{\includegraphics[width=.4\textwidth]{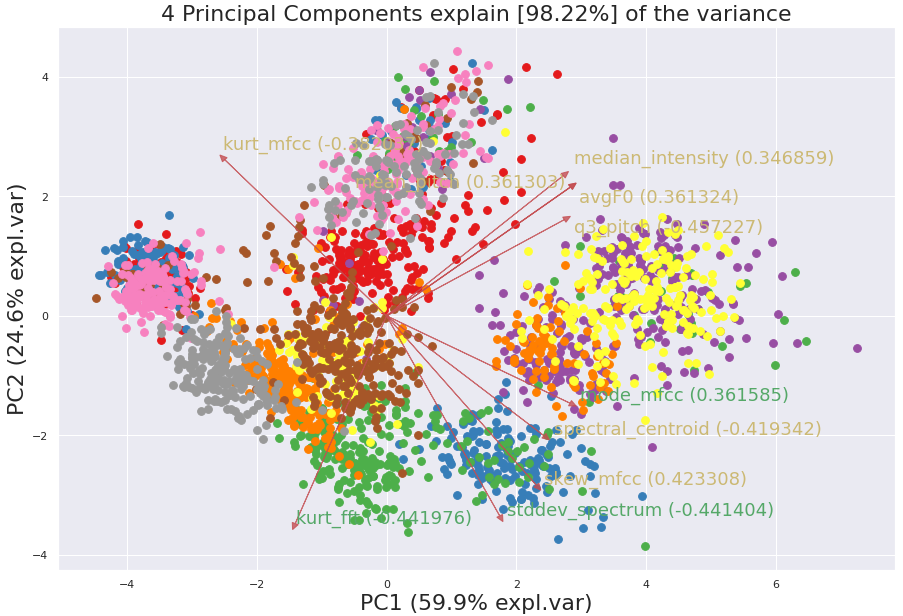}}
\caption{Biplot of TESS optimal feature combination}
\scriptsize{\textit{The length of the arrows represents the relative importance of each feature in the dataset, and the angle between the arrows represents the correlation between the features.}}
\label{fig1}
\end{figure}

The biplot in Fig. \ref{fig1} displays the data points on a 2D scatter plot, with the position of each point representing the values of the first two principal components of the optimal feature combination data. Additionally, the biplot also shows the directions and the lengths of the arrows representing the optimal features in the transformed space. Analyzing the biplot allow us to understand how the optimal features are related to the principal components as the direction of the arrow indicates the sign of the contribution (positive or negative), while the length indicates the magnitude of the contribution. This, informs us on how these features contribute to the overall variance of the data. Moreover, the cumulative explained variance 
can be added to the biplot, which indicates the percentage of the total variance in the optimal feature combination data explained by each principal component. This is helpful in determining the optimal number of principal components to retain when performing PCA to the selected feature combinations.

\subsection{Results}
\subsubsection{Comparison with SOTA methods}

\begin{table}[t]
    \centering
    \caption{Compared methods performance on TESS dataset: best results are in bold font}
    \scalemath{0.9}{
    \begin{tabular}{|l|l|l|}
        \hline
        \multicolumn{3}{|c|}{\textbf{Compared methods}}                                           \\ \hline
        \textit{\textbf{TESS dataset}}              & \textbf{Test Accuracy (\%)} & \textbf{F1-score (\%)} \\ \hline
        \textbf{Aggarwal et al. \cite{aggarwal2022two}} & 97.6  &  97 \\ \hline
        \textbf{Praseetha et al. \cite{praseetha2018deep}} &    95.8 &  NA    \\ \hline
        \textbf{Choudhary et al. \cite{choudhary2022speech}}     &   97.1 &  96 \\ \hline
        \textbf{Huang et al. \cite{https://doi.org/10.48550/arxiv.1905.08632}} &   85     &  NA \\ \hline
        \textbf{Iqbal et al. \cite{Iqbal2020MFCCAM}} &   97 &  NA  \\ \hline
        \textbf{Kapoor et al. \cite{kapoor2022fusing}}  &   97.5  &  97.4  \\ \hline
        \textbf{Krishnan et al. \cite{krishnan2021emotion}}  &   93.3  &  NA   \\ \hline
        \textbf{Dupuis et al. (HLP) \cite{dupuis2011recognition}} &    82 &  NA  \\ \hline
        \textbf{Our method} &  \textbf{98.7}  &  \textbf{98.7} \\ \hline
        \end{tabular}
        }
      \label{table3}
\end{table}

The results of our proposed method on the TESS dataset are compared with other state-of-the-art methods, as presented in Table \ref{table3}. The performance metrics of these state-of-the-art methods are taken from the original papers.
We have two main evaluation approaches for our method. Firstly, we compare our method against human-level performance (HLP) on the TESS dataset, as evaluated in \cite{dupuis2011recognition}. The authors used 56 human annotators to recognize emotions, and we use their results to evaluate our model's performance. Secondly, we compare our method against machine learning-based SER methods. As previously discussed, SER methods typically involve two main stages: feature extraction and classification. Many of the compared methods in the literature use MFCC for feature extraction \cite{Iqbal2020MFCCAM}, \cite{praseetha2018deep}, \cite{https://doi.org/10.48550/arxiv.1905.08632}, \cite{choudhary2022speech}, while some others use spectrograms \cite{kapoor2022fusing} combined with PCA \cite{aggarwal2022two} or Empirical Mode Decomposition (EMD) \cite{krishnan2021emotion}. For classification, some methods employ traditional machine learning techniques such as SVM \cite{Iqbal2020MFCCAM} or Latent Dirichlet Allocation (LDA) \cite{krishnan2021emotion}, while others use deep neural networks \cite{aggarwal2022two}, \cite{https://doi.org/10.48550/arxiv.1905.08632}, \cite{choudhary2022speech}, \cite{kapoor2022fusing}, \cite{praseetha2018deep}. 
By comparing our results to the HLP, we can see that our model is able to perform emotion recognition tasks better than humans according to the results in \cite{dupuis2011recognition}. Our method also outperformed the compared machine learning methods, achieving the highest accuracy and F1-score as shown in Table \ref{table3}.

\subsubsection{Importance of feature boosting and model explainability}

In Table \ref{table1}, we can see the performance of various machine learning models without feature boosting and model explainability on the initially computed features. It is clear from the table that the ET classifier performs the best, as it achieves 95.8\% in terms of accuracy, recall, precision and F1-score. The LGBM also performs well with 95\% accuracy. RF and GBC also have a very high accuracy at 94.6\% and 94.3\%, respectively. 
This, indicates that the initially computed features do indeed contain valuable information for the SER task. Therefor, we can conclude that the ET classifier and LGBM are the best models for this dataset, and we can use them to achieve high performance. 

\begin{figure}[t]
\centerline{\includegraphics[width=.45\textwidth]{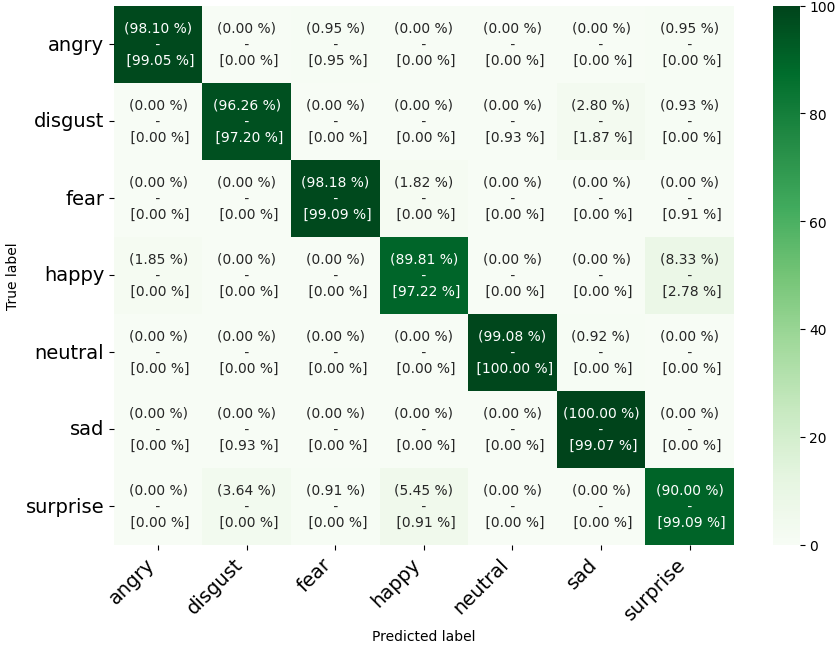}}
\caption{Confusion matrices for the Extra Trees classifier.  Values between "( )": without feature boosting and model explainability. Values between "[ ]": our framework with feature boosting and model explainability}
\label{fig3}
\end{figure}

The confusion matrix in Fig. \ref{fig3}, values between "( )", shows the results of the ET classifier. The matrix indicates that the model performs well overall, as can be seen by the high rate of correct predictions on the diagonal elements. For example, 100\% of the actual class "sad" are correctly predicted as "sad". However, there are also some misclassifications, such as when 5.45\% of the actual class "surprise" is predicted as "happy" and 8.33\% of the actual class "happy" is predicted as "surprise". This means that "happy" and "surprise" shares some acoustic characteristics.

\begin{table}[t]
\centering
\caption{Compared models on all computed features in (\%): Best results are in bold font}
\scalemath{0.9}{
\begin{tabular}{|l|l|l|l|l|}
\hline
\textit{\textbf{Model}}   & \textbf{Acc.} & \textbf{Recall}  & \textbf{Prec.}  & \textbf{F1}\\ \hline
\textbf{ET}      & \textbf{95.8}  & \textbf{95.8} & \textbf{95.8}  & \textbf{95.8}  \\ \hline
\textbf{LGBM}     &  95 &	94.9 &	95 &	95  \\ \hline
\textbf{RF}       & 94.6 &	94.6 &	94.7 &	94.5  \\ \hline
\textbf{GBC}      & 94.3 &	94.3 &	94.3 &	94.3  \\ \hline
\textbf{LDA}      & 92.9 &	93 &	93.3 & 92.9  \\ \hline
\textbf{DT}     & 92.9 &	92.1 &	92.3 &	92.1  \\ \hline
\textbf{QDA}   & 83.2 &	83.3 &	85.7 &	82  \\ \hline
\end{tabular}
}
\label{table1}
\end{table}

\begin{table}[t]
\centering
\caption{Compared models on the constructed dataset using feature boosting and model explainability in (\%): Best results are in bold font}
\scalemath{0.9}{
\begin{tabular}{|l|l|l|l|l|}
\hline
\textit{\textbf{Model}}   & \textbf{Acc.} & \textbf{Recall}  & \textbf{Prec.}  & \textbf{F1}\\ \hline

\textbf{ET}		& \textbf{98.7}	& \textbf{98.6}	& \textbf{98.7}	& \textbf{98.7} \\ \hline
\textbf{LGBM}	& 	95.9 &  96 &  96.3 &  96 \\ \hline
\textbf{RF}		& 95  &  95 &	95 &  95\\ \hline
\textbf{LDA}	& 94.5 &  94.5 &	94.6 &  94.5 \\ \hline
\textbf{GBC}   & 93.6 &  93.6 &	93.7 &  93.7 \\ \hline
\textbf{QDA}	& 92.6 &  92.7 &	92.7 &  92.6 \\ \hline
\textbf{DT}		& 90.3 &  90.3 &	90.6 &  90.2 \\ \hline
\end{tabular}
}
\label{table2}
\end{table}

Table \ref{table2} compares the performance of the same machine learning models using the boosted feature set and the incorporation of model explainability feedback. 
The best performing model in terms of accuracy is the ET classifier with an accuracy of 98.7\% and F1-score of 98.7\%. The second best performing model is LGBM with an accuracy and F1-score of 95.9\% and 96\%, respectively. RF also has relatively high accuracy of 95\%, and good scores for the other evaluation metrics. 
Some other models, have lower accuracy and less favorable scores for the other evaluation metrics. Few of these low-performing models appear to lose their effectiveness when feature boosting and model explainability are employed. This may be due to their need for more complex datasets in order to achieve reasonable output, resulting in increased computational complexity. This suggests that these models may be less suitable for SERs than the models that perform well under the same conditions. In summary, the ET classifier and LGBM are the best performing models on this dataset with high accuracy and F1-score. 

The confusion matrix in Fig. \ref{fig3}, values between "[ ]", shows the results of the ET classifier, which performs well overall with a high number of correct predictions on the diagonal elements for each emotion. For example, 100\% of the actual class "neutral" are correctly predicted as "neural" and 99.09\% of the actual class "fear" are correctly predicted as "fear". However, there are also some misclassifications, such as 1.87\% of "disgust" being predicted as "sad" and 2.78\% of "happy" being predicted as "surprise". This means that "happy" shares characteristics with "surprise" as well as "disgust" with "sad".

Furthermore, the process of hyperparameters tuning can help to optimize the performance of a model by finding the optimal values for the hyperparameters that control the model's complexity and generalization performance, such as the case of our compared models. We achieve this by using random grid search technique, which involves training the model using a range of hyperparameter values and evaluating the performance of each model using cross-validation set. The best set of hyperparameters can then be selected based on the model's performance on the validation set.

To sum up, the use of feature boosting and model explainability to refine and boost feature selection results in a significant increase and robustness in the performance of the models, which can be attributed to several factors. One of the main reasons is that PCA is able to reduce the dimensionality of the data, which can help to remove noise and redundancy. This induces a more compact and informative representation of the data that is better suited for the machine learning models. Additionally, the explainability module drives the feedback loop by identifying the most important combinations of features that are driving the classification. This allows us to iteratively boost the feature selection process, taking into account new insights and understanding gained through XAI techniques. This iterative process leads to even better representation of the speech signal via the most important features for the SER task, resulting in improved model generalization.

\section{Discussion}
In this work, the objective is to extract the most relevant features for accurately detecting emotions in speech. We noticed a lack of consensus in the existing literature regarding a clear set of features that effectively capture emotional information from speech signals. This knowledge gap highlights the importance of our research in addressing this issue and providing valuable insights into feature selection for SER. Through this study, we make significant contributions to the field by exploring and identifying the features that play a crucial role in detecting and distinguishing emotional states in speech. 

Our analysis focused on identifying features with high discriminative power and informativeness for differentiating between various emotional categories and how can we explain and interpret that. We employ a rigorous feature selection process to identify the most relevant features. By using an iterative feature boosting technique supported by the explainabilty module, we aim to enhance the discriminative power of the selected features and the SER's overall performance. 

One of the key features that emerges as highly relevant in our study is MFCC, which capture the spectral characteristics of speech and have been widely used in speech analysis tasks. MFCCs are known to effectively capture the distinctive characteristics and spectral variations associated with different patterns. Another important feature that we find valuable for SER is pitch or fundamental frequency (F0). Variations in pitch convey important emotional cues and can help discriminate between different emotional states. Analyzing pitch-related features such as pitch contour, pitch range, or pitch dynamics can provide valuable information for emotion classification. Additionally, we find that energy and intensity measures play a significant role in capturing emotional intensity and arousal. These features reflect the overall energy distribution and loudness of speech, which are closely related to emotional expressiveness. Furthermore, temporal features such as speech rate or duration demonstrated relevance in capturing the temporal patterns and dynamics of emotional speech. The rate at which speech is produced and the duration of specific speech segments can provide important cues for SER. 

One limitation of our study is that it was tested solely on the TESS dataset. While the results obtained from this dataset are promising, it is important to validate our approach on multiple datasets and in real-world scenarios to ensure the generalizability of our findings. Validation on diverse datasets would provide a more comprehensive assessment of the effectiveness of our feature boosting approach in different contexts and with different speech samples. It would enable us to evaluate the robustness and reliability of the selected features across various data sources, potentially uncovering any dataset-specific biases or limitations. Furthermore, real-world scenarios often present additional challenges, such as varying recording conditions, speaker characteristics, and noise levels. These factors can impact the performance of the SER system and the relevance of the selected features. 

However, it is worth noting that our research provides a comprehensive analysis of the feature selection process and highlights the rationale behind selecting specific features. By presenting these findings, we contribute to the development of a standardized feature set for SER, which can serve as a foundation for future research in the field. This standardized set of features will enable researchers to focus on these key features when designing and implementing robust SERs, ultimately enhancing the accuracy and effectiveness of trustworthy emotion detection in real-world applications.

\section{Conclusion}
In this study, we presented a new approach for supervised SER based on acoustic features of the voice and their statistical characteristics values. Our work highlights the importance of feature selection and the role of explainability in improving the accuracy of SER. Our method involves several steps including computing speech features, selecting the optimal feature combinations, boosting the final feature subset, and applying various machine learning models to evaluate their performance and fine-tune the best performing model. Additionally, we incorporate XAI into SER to create a system that is more understandable, to boost feature selection process via the feedback loop. To the best of our knowledge, this is the first work incorporating model explainability into an SER framework. Our work provides a comprehensive SER approach that aims to balance the benefits of advanced machine learning techniques with the need for transparency and comprehensibility. Our future work aims to investigate feature boosting within deep learning frameworks. Moreover, we aim to generalize our feature boosting approach for other classification problems, addressing the feature relevance and high dimensionality problems.

\bibliographystyle{ieeetr}
\bibliography{references}

\end{document}